\documentclass{pasj00}

\SetRunningHead{Astronomical Society of Japan}{Usage of \texttt{pasj00.cls}}
\usepackage{times}

\begin{document}

\title{Observational Report on the Classical Nova KT Eridani}
\SetRunningHead{K. Imamura and K. Tanabe}{Observational Report on the Classical Nova KT Eri}

\author{Kazuyoshi \textsc{Imamura}$^1$ and Kenji \textsc{Tanabe}$^2$
        }
\affil{$^1$Department of Mathematical and Environmental System Science, Faculty of Informatics,\\
       Okayama University of Science, 1-1 Ridai-cho, kita-ku, Okayama 700-0005}
\email{imako@pc.117.cx}
\affil{$^2$Department of Biosphere-Geosphere Science, Faculty of Biosphere-Geosphere Science,\\
       Okayama University of Science, 1-1 Ridai-cho, kita-ku, Okayama 700-0005}
\email{tanabe@big.ous.ac.jp}

\Received{2012 January 30}
\Accepted{2012 May 21}


\maketitle

\begin{abstract}
   
  A report on the spectroscopic and multi-color photometric observations 
of high galactic latitude classical nova KT Eridani (Nova Eridani 2009) is presented. 
After 12.2 days from maximum light, broad and prominent emission lines of 
Balmer series, He I, He II, N II, N III and O I can be seen on the spectra. 
The FWHM of H${\alpha}$ line yields an expansion velocity of approximately 
3400 km s$^{-1}$. After 279.4 days from maximum light, we can see prominent 
emission lines of He II and [O III] on the spectrum. 
Among them, [O III] (4959, 5007) lines show multiple peaks. 
From the obtained light curve, KT Eri is classified to be a very fast nova, 
with a decline rate by two magnitude of $6.2 \pm 0.3$ days and 
three of $14.3 \pm 0.7$ days. 
We tried to estimate the absolute magnitude ($M_V$) using the Maximum 
Magnitude versus Rate of Decline relationship and distance of KT Eri. 
The calculated $M_V$ is approximately $-9$. 
Accordingly, the distance and galactic height are approximately 7 kpc and 
4 kpc, respectively. Hence, KT Eri is concluded to be located outside of the 
galactic disk.

\end{abstract}

\section{Introduction}

   Galactic novae, including both classical and recurrent novae, had been 
detected along the galactic equator and concentrated strongly to the galactic 
center (see \cite{war2008} for a review). 
This is due to the possibility of nova eruption which is directly proportional 
to the number of Population I stars. Figure 1 is a nova map detected by the 
year of 2010 in galactic coordinates. From this figure, the portion of novae 
with $|b|>20^{\circ}$ is at most about 3\%.

   Classical novae (CNe, hereafter), discriminated from the recurrent novae (RNe) 
by their single eruption record, are one category of the cataclysmic variable 
stars that are close binary system of white dwarf and normal star.
This binary property was based on the photoelectric observation and the radial 
velocity observation by \citet{wal1954} and \citet{kra1958} during DQ Her in 
its quiescence, respectively. 
Based on the accumulated light curves, nova classification  
using the speed class was proposed by \citet{pay1957} and \citet{due1981}. 
On the other hand, spectral evolution and 
classification of CNe had been performed by Williams (1990, 1992).
     
   Nova eruption is caused by the thermonuclear runaway reaction on a surface 
of the white dwarf (\cite{gal1978} and \cite{sta1989}). 
Mass transfer containing rich hydrogen gas from the secondary star with high 
temperature and pressure due to the strong surface gravity of the white dwarf 
leads to thermonuclear runaway.

   KT Eridani (Nova Eridani 2009) was detected  by a Japanese skillful observer 
K. Itagaki on 2009 November 25.536 (UT) at 8.1 magnitude (\cite{yam2009}). 
He reported the position of this object as R.A. $=$ \timeform{4h47m54s.21}, 
Dec. $=$ \timeform{-10D10'43".1} (the equinox 2000.0). 
At first this object had a possibilty of WZ Sge-type dwarf nova because of its 
smaller rising amplitude than typical novae (by T. Kato in \textit{venet-alert 11692}
\footnote{http://ooruri.kusastro.kyoto-u.ac.jp/mailarchive/vsnet-alert/11692}), 
but afterwards the star was confirmed to be a nova explosion by spectroscopic 
observations (\cite{yam2009}). This star is historically the first 
classical nova ever detected at the constellation of Eridanus. 
As the location of KT Eri is quite exceptional with high galactic latitude and 
anti-galactic center ($l=208^{\circ}, b=-32^{\circ}$), 
this nova is supposed to have something different from ordinary CNe.

  Apart from the optical observations, KT Eri was detected at both radio and X--ray
(\cite{OBr2010}, \cite{Bod2010}).
The radio observations indicate that KT Eri had been detected during its rising stage
of the radio light curve. After 55.4 days from maximum light, 
\textit{Swift} XRT detected a super soft source (SSS). 
After 65.6 days from the maximum, the SSS was softened drastically.

   The main purpose of this paper is to present our optical observational report of 
classical nova KT Eri from the earliest time to the later stage (more than half a year)
of its eruption. 
In addition we remark the peculiarity of this nova from the point of view of its 
galactic location.
In \S2, we present the spectroscopic and multi-color photometric observations. 
\S3 gives results. \S4 is for the discussions. Summary is stated in \S5.
     
    \begin{figure*}
      \begin{center}
        \FigureFile(110mm,110mm){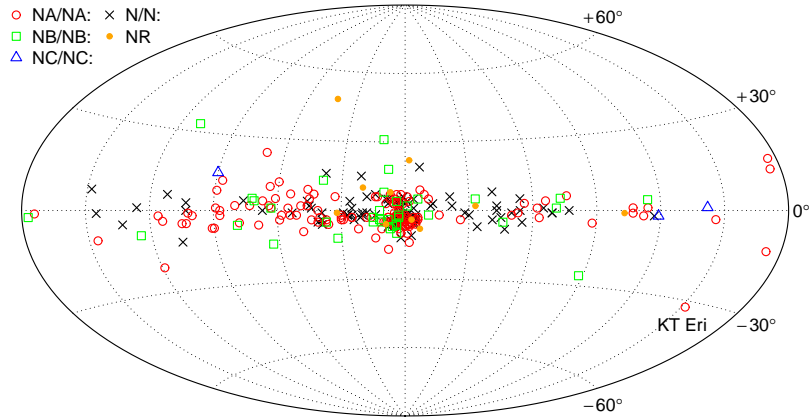}
      \end{center}
     \caption{The distribution of classical and recurrent novae in galactic coordinates (an Aitoff projection) 
              depicted by the present author (KI). 
              The data are from CV catalog (\cite{dow2005}) and IAU circulars (2006-2010 CNe).
              Red circles, green squares, blue triangles, black crosses and orange filled circles mark are 
              fast novae (NA/NA:), slow novae (NB/NB:), extremely slow novae (NC/NC:), uncategorized novae (N/N:)
              and recurrent novae (NR), respectively.}
    \end{figure*}

\section{Observations}

 \subsection{Spectroscopy}
   Our observational systems of spectroscopy are shown in Table 1. 
The first one, the system of OUS (Okayama Univ. of Sci.) observatory is a 
combination of DSS-7 (SBIG production) spectrometer and ST-402 (SBIG) CCD 
camera installed on Celestron 28cm (F/10) Schmidt-Cassegrain telescope. 
The second system of BAO (Bisei Astronomical Observatory) is a combination of 
low-resolution spectrometer and DU-440BV (ANDOR) CCD camera installed on 101cm 
(F/12) Classical Cassegrain Telescope. The spectrometer's resolution 
($R = \lambda / \Delta\lambda$) of OUS and BAO is approximately 400 and 1000, 
respectively. Wavelength calibrations of OUS and BAO were made by  
Hydrogen--Helium lamp and Iron--Neon lamp, respectively. We have performed the 
spectroscopic observations from 2009 November 26 to 2010 January 26 at OUS 
observatory. Additionally, we observed the nova on 2010 August 20 at BAO. 
The total number of observation is 34 nights. 
Table 2 is log of spectroscopic observations.

 \subsection{Multi-color photometry}

\begin{table}
  \caption{Instruments of spectroscopy.}
  \begin{center}
 \begin{tabular}{lccc} \hline
  Observatory\footnotemark[$*$] & Telescope\footnotemark[$\dagger$] & $R$\footnotemark[$\ddagger$] & wavelength (\AA) \\ \hline
   OUS            &  28cm SCT      & 400           & 4200 -- 8200       \\
   BAO            &  101cm CCT     & 1000          & 4000 -- 8000       \\
 \hline
 \end{tabular}
 \end{center}
 \footnotesize {\bf Note.\\}
 \noindent
 \footnotemark[$*$]OUS (Okayama Univ. of Sci.), BAO (Bisei Astronomical Observatory).
 \par\noindent
 \footnotemark[$\dagger$]SCT (Schmidt-Cassegrain telescope), CCT (Classical Cassegrain Telescope).
 \par\noindent
 \footnotemark[$\ddagger$]$R = \lambda / \Delta\lambda$ at 6000 \AA. 
\end{table}

\begin{table}
  \caption{Log of spectroscopic observations.}
  \begin{center}
 \begin{tabular}{crr} \hline
  JD\footnotemark[$*$]  &  Date (UT)             & Observatory \\ \hline
  5162.12               &  2009 November 26.62   & OUS         \\
  5164.16               &                28.66   & OUS         \\
  5165.21               &                29.71   & OUS         \\
  5166.17               &                30.67   & OUS         \\
  5167.13               &       December  1.63   & OUS         \\
  5168.13               &                 2.63   & OUS         \\
  5169.13               &                 3.63   & OUS         \\
  5170.11               &                 4.61   & OUS         \\
  5172.12               &                 6.62   & OUS         \\
  5174.11               &                 8.61   & OUS         \\
  5175.11               &                 9.61   & OUS         \\
  5183.13               &                17.63   & OUS         \\
  5184.07               &                18.57   & OUS         \\
  5185.08               &                19.58   & OUS         \\
  5186.07               &                20.57   & OUS         \\
  5187.12               &                21.62   & OUS         \\
  5188.06               &                22.56   & OUS         \\
  5195.04               &                29.54   & OUS         \\
  5197.03               &                31.53   & OUS         \\
  5200.08               &  2010 January   3.58   & OUS         \\
  5202.05               &                 5.55   & OUS         \\
  5203.03               &                 6.53   & OUS         \\
  5204.02               &                 7.52   & OUS         \\
  5205.10               &                 8.60   & OUS         \\
  5209.00               &                12.50   & OUS         \\
  5210.04               &                13.54   & OUS         \\
  5211.01               &                14.51   & OUS         \\
  5213.04               &                16.54   & OUS         \\
  5214.00               &                17.50   & OUS         \\
  5215.00               &                18.50   & OUS         \\
  5220.04               &                23.54   & OUS         \\
  5222.01               &                25.51   & OUS         \\
  5223.01               &                26.51   & OUS         \\
  5429.29               &       August   20.79   & BAO         \\

 \hline
 \end{tabular}
 \end{center}
 \footnotesize {\bf Note.}
 \noindent 
 \footnotemark[$*$] JD$-2450000$.
\end{table}

      Our system of multi-color photometry is a combination of ST-7E (SBIG) 
CCD camera accompanied with $B, V, R_c$ and Str\"{o}mgren $y$ filter attached 
to Celestron 23.5cm (F/6.3) Schmidt-Cassegrain telescope. 
We have performed photometric observations from 2009 November 26 to 2010 
March 19 and from 2010 August 5 to December 10 at OUS observatory. 
The total number of observation is 79 nights.

\section{Results}

  \subsection{Spectroscopic observations}
  
    \begin{figure*}
      \begin{center}
        \FigureFile(120mm,120mm){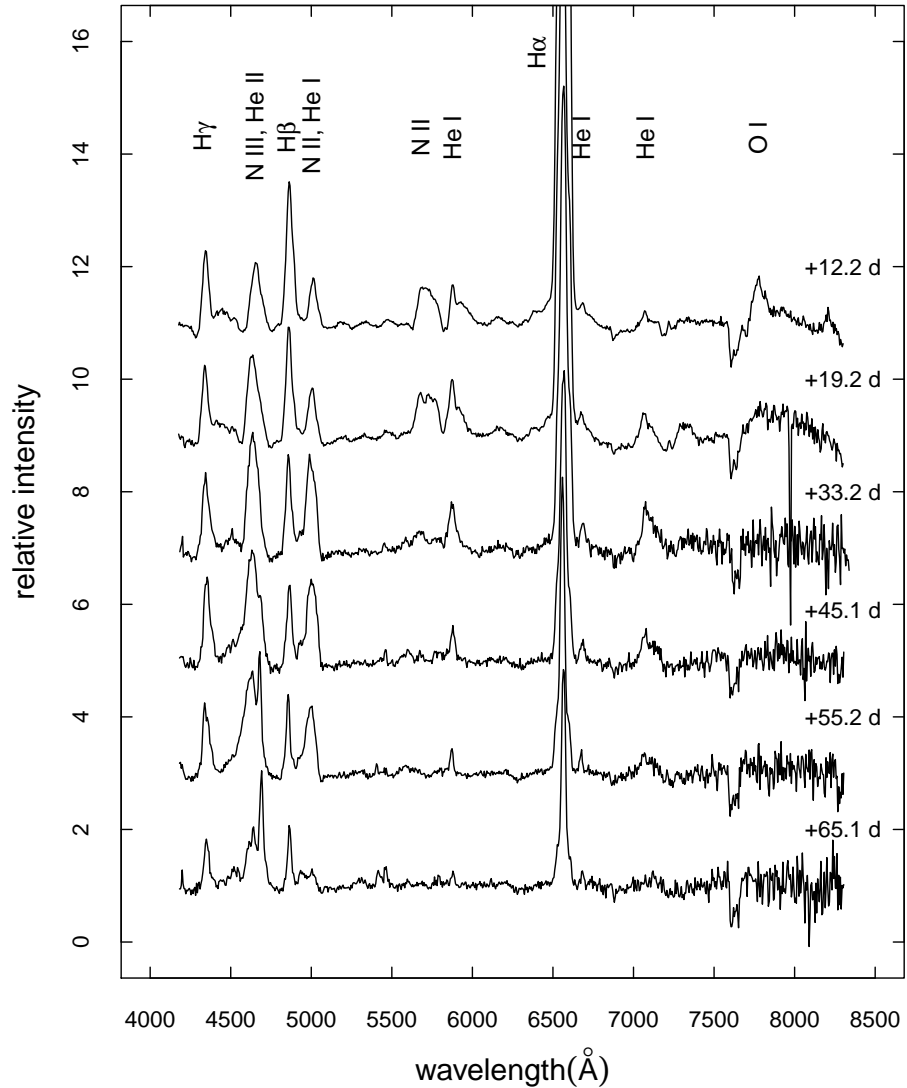}
      \end{center}
     \caption{Our representative spectra of KT Eri. 
              The numerical values on the right edge are the elapsed days from the maximum light.
              Each of these continuum is normalized as unity. These are obtained at OUS observatory.}
    \end{figure*}
  
   Figure 2 shows representative spectra of KT Eri at OUS (R$\sim400$). 
Broad and prominent emission lines of Balmer series (H$\alpha$, H$\beta$, H$\gamma$), 
He I (5016, 5876, 6678, 7065), He II (4686), N II (5001, 5679), N III (4640) 
and O I (7773) are seen on the spectra. 
Identification of these emission lines is based on those works by 
\citet{wil1992} and \citet{wil1994}. The FWHM (Full Width at Half Maximum) 
of H$\alpha$ line is approximately 3400 km s$^{-1}$ on the first observational night 
($+12.2$ days from maximum light; Fig. 2). 
At first, He II (4686) and N III (4640) lines are blended. Then, after 55 days from 
maximum light, these lines are split. Moreover He II (4686) become stronger and 
narrower than before. According to the spectra, this nova is classified as a 
He/N nova (\cite{wil1992}).  

    \begin{figure*}[t]
      \begin{center}
        \FigureFile(150mm,150mm){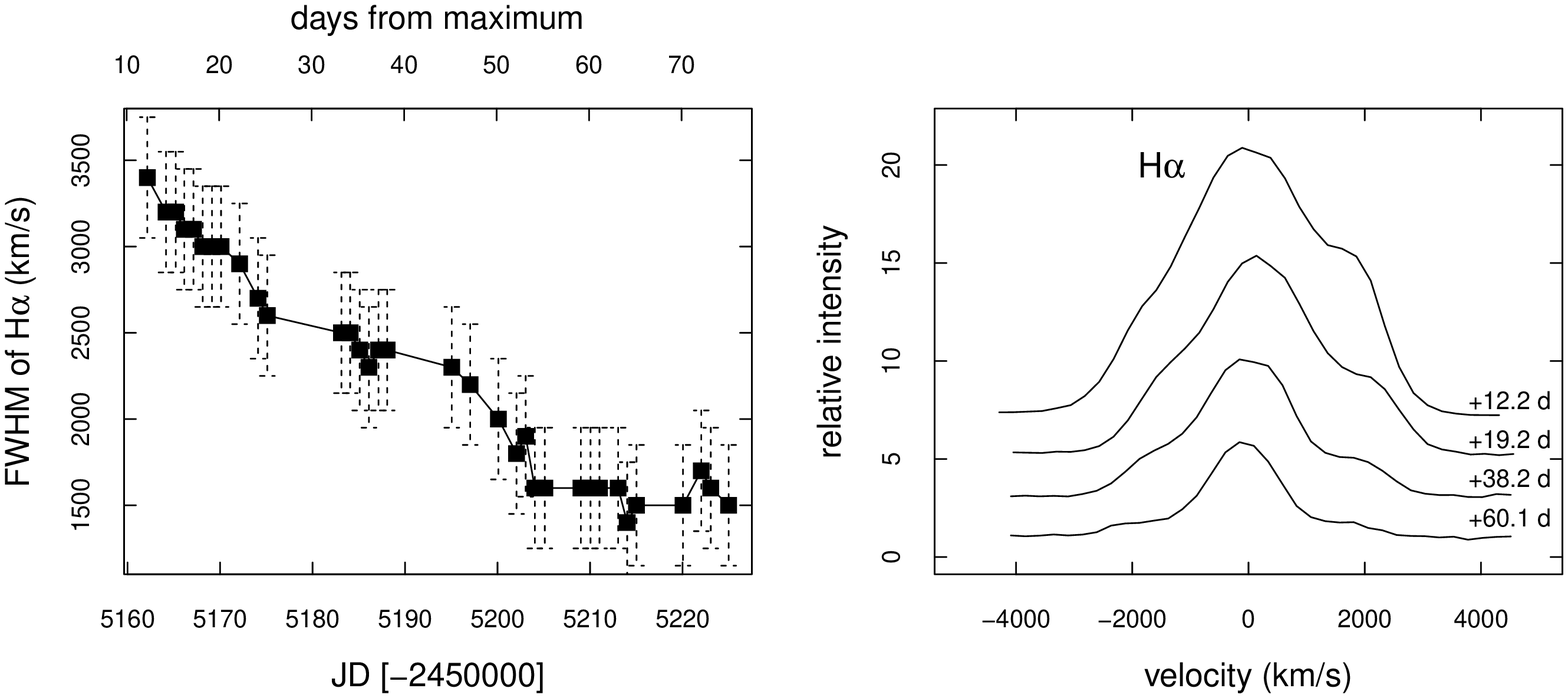}
      \end{center}
     \caption{The left one is the FWHM of H$\alpha$'s temporal variation. 
              The right one is temporal variation of H$\alpha$ profiles.}

    \begin{center}
        \FigureFile(160mm,160mm){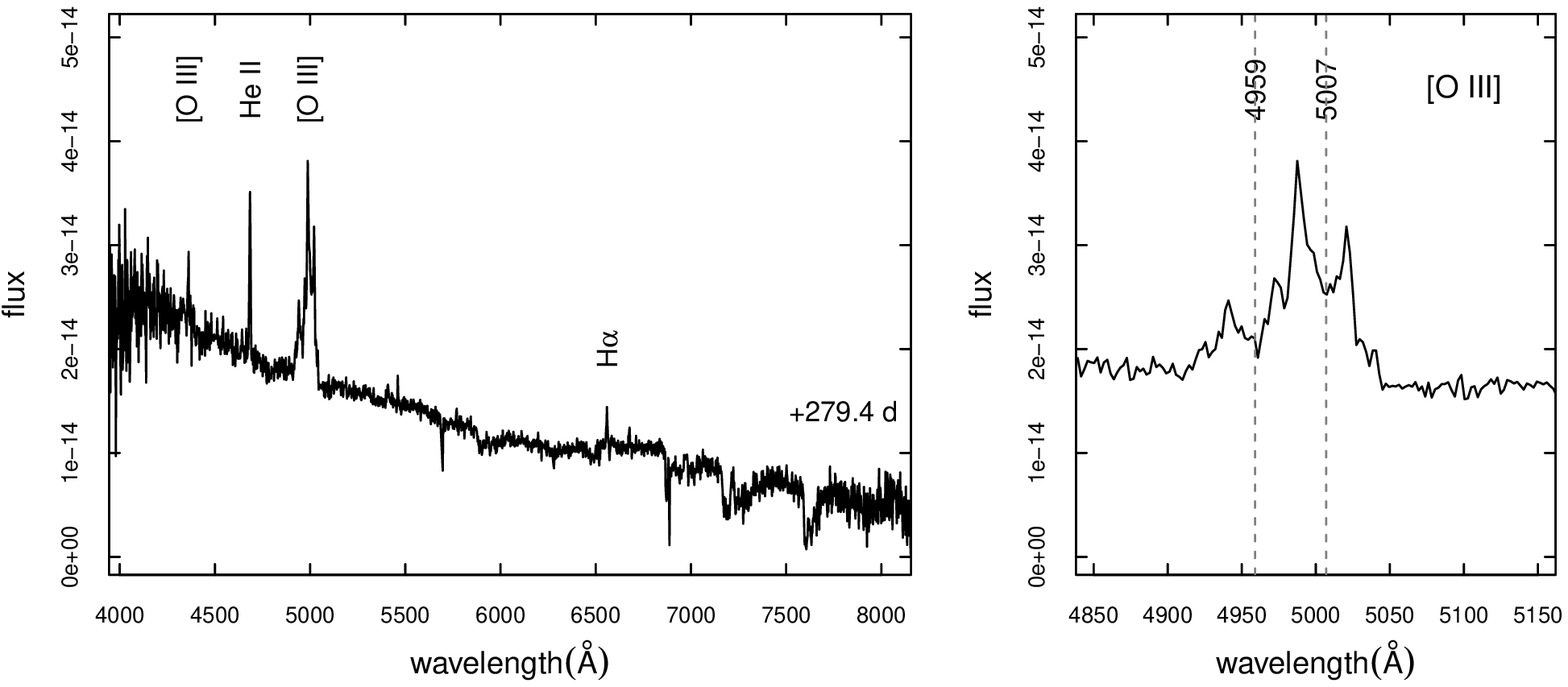}
    \end{center}  
     \caption{Spectrum of KT Eri in nebular phase (after 279.4 days from maximum light) at BAO. 
              The right one is enlargement of [O III] (4959, 5007) region.
              The unit of vertical axis is erg s$^{-1}$ cm$^{-2}$ \AA $^{-1}$.}
    \end{figure*}

  Figure 3 (left) shows the FWHM of H$\alpha$'s temporal variation. 
The decline rate of FWHM is $\sim 30$ km s$^{-1}$ per day and  
down to 1500 km s$^{-1}$ about 70 days after the maximum light. 
Figure 3 (right) is a temporal change of H$\alpha$ profiles. 
It shows asymmetric profiles at the earlier stages. 
Such a tendency suggests the existence of the non-spherical 
expanding gas shell.

  Figure 4 is a spectrum of KT Eri in nebular phase at BAO (R$\sim$1000). 
We can see prominent emission lines of He II (4686) and 
[O III] (4363, 4959, 5007) on the spectrum. [O III] (4959, 5007) lines 
show multiple peaks (at least five). 
These peaks suggest the existence of the multi-ejecta. 
On the other hand, H$\alpha$ line seems to be very weak. 
Such the profile of [O III] is seen in the coronal phase of V444 Sct 
(Nova Sct 1991; \cite{wil1994}). Speed class of V444 Sct is 
\textit{very fast}. 
Hence KT Eri seems to have common features of V444 Sct.

  \subsection{Multi-color photometric observations}
  
    \begin{figure*}
      \begin{center}
        \FigureFile(140mm,140mm){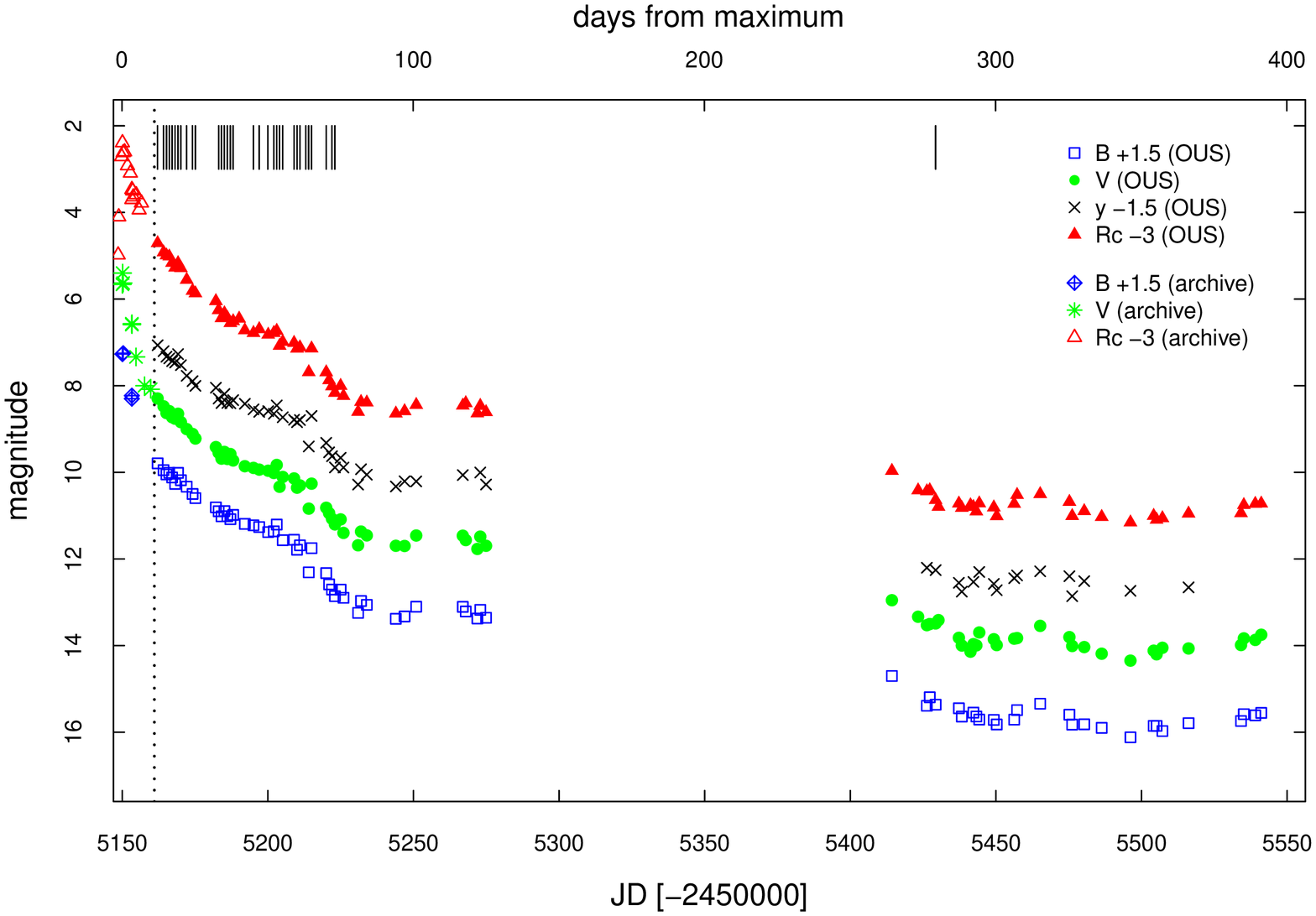}
      \end{center}
     \caption{The result of multi-color photometric observations. 
              The data before the discovery are archival one by ASAS, Pi of the sky and VSOLJ.
              Dashed line is discovery date. Tick marks indicate the corresponding epochs to 
              the date of our spectroscopic observations.}
    \end{figure*}

  Figure 5 is a result of our multi-color ($B, V, R_c, y$) 
photometric observations. The data before the discovery are 
based on archival ones by \textit{Pi of the Sky} 
\footnote{http://grb.fuw.edu.pl/}, \textit{All Sky Automated 
Survey} \footnote{http://www.astrouw.edu.pl/asas/, \citet{asas}} (collected by T. Kato in 
\textit{vsnet-alert 11697}
\footnote{http://ooruri.kusastro.kyoto-u.ac.jp/mailarchive/vsnet-alert/11697})
and Japanese amateur astronomer with Digital Single Lens Reflex camera 
(\cite{Oot2009}; also posted in VSOLJ (Variable Star Observers League in Japan)).
From these data, the maximum brightness is supposed 
to be 5.4 $V$ magnitude. In its early decline phase, 
the magnitude change shows a rapid fading by 0.32 $V$ magnitude per day. 
From the derived parameters maximum date $t_0$ is 2009 
November $14.4 \pm 0.2$ UT (2455149.9 JD). 
The decline time $t_2$ and $t_3$ are $6.2 \pm 0.3$ days 
and $14.3 \pm 0.7$ days, respectively. 
According to this result, the speed class is thought to 
be \textit{very fast} ($t_2 < 10$; \cite{pay1957}). 
The statistical relationship between $t_2$ and $t_3$ 
($t_3 \approx 2.75 t_2^{0.88}$; \cite{war1995}) agrees well with the observed values. 

  Figure 6 is our result of the variations of color index ($B-V$ and $V-R_c$). 
Each color index became bluer. The $B-V$ shows bluest around 2455187 JD 
(about 37 days from maximum light).

\section{Discussions}

\subsection{Spectral evolution}
  According to \citet{wil1992}, novae showing stronger lines of He and
N have larger expansion velocities and higher level of ionization.
Such a type of nova also shows very rapid evolution.
Hence the spectral class and speed class of KT Eri is thought to be 
a He/N nova.

  The spectral evolution of KT Eri is similar to that of rapidly evolving 
CNe or RNe. As mentioned in 3.1, from figure 2 we can see that the blended 
lines of He II (4686) and N III (4640) are split, and He II narrow line emerges 
and grows. Such a behavior had been observed in Nova LMC 1990 No. 1 
(CN of He/N), V444 Sct (CN of Fe II), V394 CrA (RN of He/N), 
U Sco (RN of He/N), V745 Sco (RN of He/N) by Williams et al. (1991, 1994) 
and Sekiguchi et al. (1988, 1989). Sekiguchi et al. (1988, 1989) 
suggests that He II narrow line in V394 CrA and U Sco during the outburst 
are attributed to an accretion disc. On the other hand, 
\citet{diaz2010} in U Sco during the 2010 outburst proposes the following two ideas: 
(1) the reionization of circumbinary gas from previous outburst or 
(2) chromospheric emission from X--ray illumination of the companion 
by the shrinking nova photosphere. Both X--ray (Super Soft Source) and He II (4686) 
become strong after around 60 days from maximum light. 
It is thought that its evolution is correlated with growth of He II (4686), 
and this phenomenon supports the latter idea of \citet{diaz2010}.

\subsection{Distance and galactic location}

  We have tried to estimate the absolute magnitude at maximum using four Maximum 
Magnitude versus Rate of Decline (MMRD) relations to derive the distance to KT Eri.
The color excess of KT Eri estimated by \citet{rag2009} is  
$E(B-V) \sim 0.08$, which is based on \citet{mun1997}.
Table 3 shows the obtained results by using various parameters. 
Accordingly, the estimated absolute magnitude at maximum is approximately $-9$. 
The resultant distance is $6.6 \pm 0.8$ kpc at weight average. 

  Using the above results, we can discuss the spatial location of KT Eri. 
Taking into account that the galactic latitude of KT Eri is $-32^{\circ}$ and 
the distance is approximately 7 kpc, its galactic height is 4 kpc. 
So KT Eri is thought to be outside of the galactic thick disk. 
If KT Eri is inside the thick disk, the apparent magnitude of maximum becomes 
brighter than 3 mag. This value seems to be inconsistent with the observations. 
It was also pointed out by Solar Mass Ejection Imager (\cite{hou2010}) that 
the maximum is about 5 mag. 

  According to \citet{del1992} and \citet{del1998}, typical fast ($t_2 \le 12$) 
and He/N novae are concentrated at low height above the galactic plane. 
Taking account that these novae are located at $z < 100$ pc and are member of 
Population I star, the location of KT Eri is quite exceptional.

     \begin{figure}
      \begin{center}
        \FigureFile(80mm,80mm){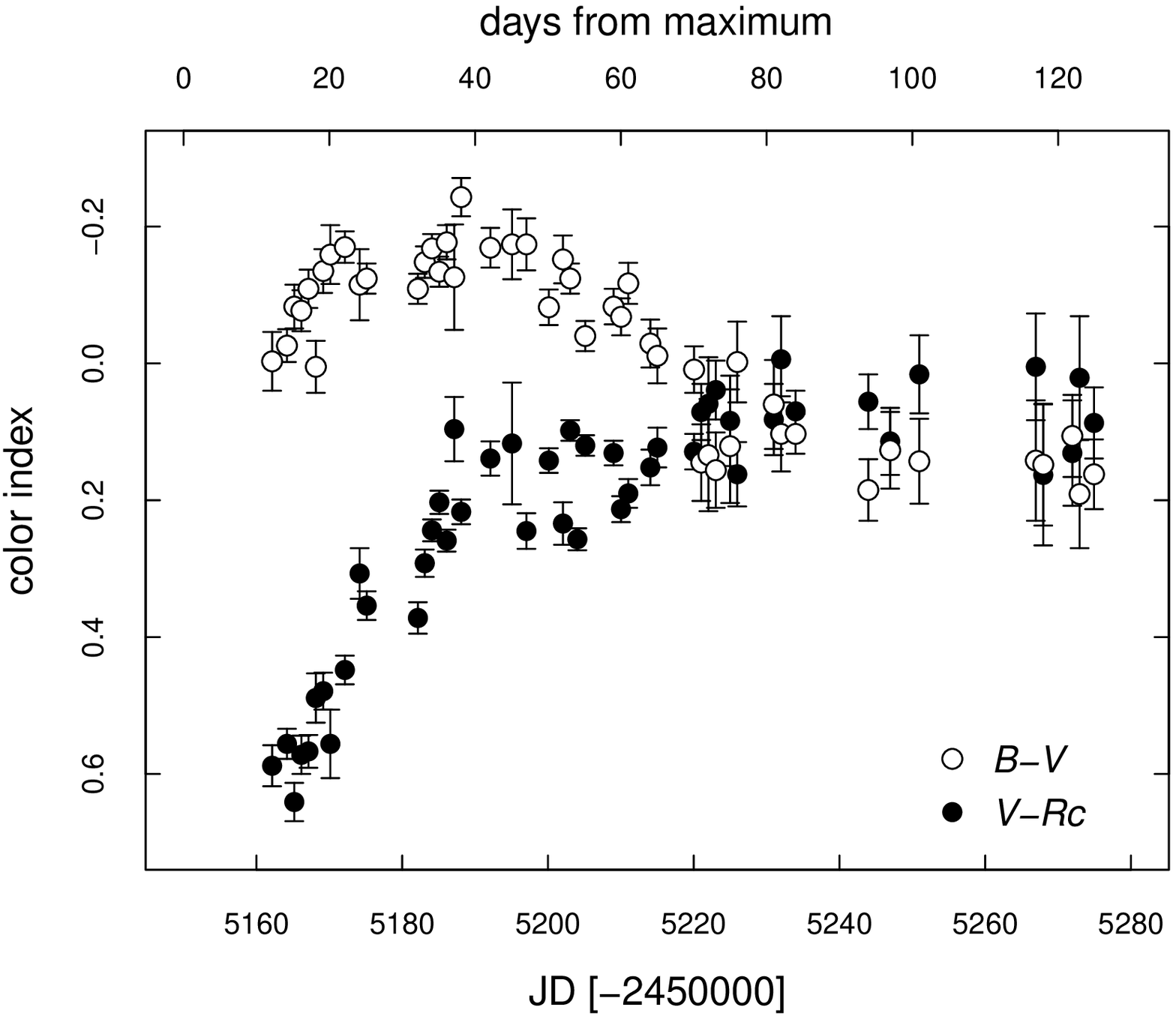}
      \end{center}
     \caption{The result of variations of color index ($B-V$ and $V-R_c$).}
    \end{figure}

 \begin{table}
  \caption{Absolute magnitude at maximum of KT Eri estimated using MMRD calibrations for $t_2$.}
  \begin{center}
 \begin{tabular}{lccc} \hline
  MMRD & $M_V$ & $d$ (kpc) \\ \hline
  \citet{coh1988}      & $-8.79 \pm 0.53$ & $6.28 \pm 1.55$ \\
  \citet{cap1989}      & $-8.88 \pm 0.61$ & $6.55 \pm 1.87$ \\
  \citet{del1995}      & $-8.86 \pm 0.41$ & $6.49 \pm 1.23$ \\
  \citet{dow2000}      & $-9.30 \pm 0.69$ & $7.94 \pm 2.57$ \\
 \hline
 \end{tabular}
 \end{center}
 \end{table}

\subsection{Absolute magnitude at minimum}
  The estimated distance and the apparent magnitude of a possible progenitor 
will give the pre-nova visual absolute magnitude of KT Eri. 
There exists a star in the Guide Star Catalog at the exact position of KT Eri. 
If this star (GSC 5325.1837 $\sim 14.8$ mag) is its true progenitor, 
its absolute magnitude is approximately $0.4$. Taking into account that the 
absolute magnitude of CNe at minimum is $4.4$ (\cite{war1987}), 
this result is much brighter by $4$ magnitude. It is important to compare 
KT Eri with RNe. Table 4 shows the properties of RNe and KT Eri. 
KT Eri has parameters similar to those of RNe. 
Some of the recurrent nova systems contain giant secondary. 
Therefore absolute magnitude at minimum is brighter than CNe. 
If the secondary of KT Eri is a giant, 
its brightness in quiescence can naturally be explained. 

 \begin{table}
  \caption{Comparison of KT Eri with RNe including giant secondary. 
           RNe data from \citet{sch2010}.}
 \begin{center}
 \begin{tabular}{lccc} \hline
  star      & $t_2$(d)  & $M_V$(max) & $M_V$(min)  \\ \hline
  T CrB     & $4.0$     & $-7.3$     & $0.3$       \\
  RS Oph    & $6.8$     & $-10.6$    & $-4.1$      \\
  V3890 Sgr & $6.4$     & $-8.6$     & $-1.2$      \\
  V745 Sco  & $6.2$     & $-8.0$     & $1.3$       \\
  \bf KT Eri& $6.2$     & $-9.0$     & $0.4$       \\
  \hline
 \end{tabular}
 \end{center}
 \end{table}

\subsection{2MASS data at quiescence}

  We have plotted the color-color diagram of $(J-H)_0$ vs $(H-K_s)_0$ for 
KT Eri (Table 5) and several RNe (Fig. 7). 
The other RNe (V394 CrA, CI Aql, IM Nor) are not plotted because 
there is no data ($J, H, K_s$) at quiescence. Data of CI Aql and IM Nor 
by \citet{darn2012} give an upper limits. 
The color of RNe containing M-giant secondary 
(T CrB, RS Oph, V745 Sco and V3890 Sgr) is comparatively red.
On the other hand, the color of KT Eri is bluer which is rather similar to 
T Pyx and U Sco. It suggests that an evolution of secondary star of KT Eri 
does not so advanced as RNe with M-giant secondary while this star evolves 
more than a secondary of T Pyx and U Sco. 
On the other hand, the periodicity and the quiescence magnitude of KT Eri are 
suggested that the secondary star is evolved and likely in, or ascending, 
the Red Giant Branch by \citet{jur2012}. Our results support their opinion.

  \begin{table}
  \caption{The progenitor of KT Eri at 2MASS Point Source Catalog.
           $A$ is the extinction in the $J$, $H$ and $K_s$ filter which calculated 
           with \citet{car1989}'s relation as $A_V = 0.2$. 
           }
  \begin{center}
 \begin{tabular}{lcc} \hline
   filter  & magnitude  & $A$      \\ \hline
     $J$   & 14.62      & 0.06     \\
     $H$   & 14.15      & 0.04     \\
     $K_s$ & 14.09      & 0.03     \\
 \hline
 \end{tabular}
 \end{center}
 \end{table}

    \begin{figure}
      \begin{center}
        \FigureFile(70mm,70mm){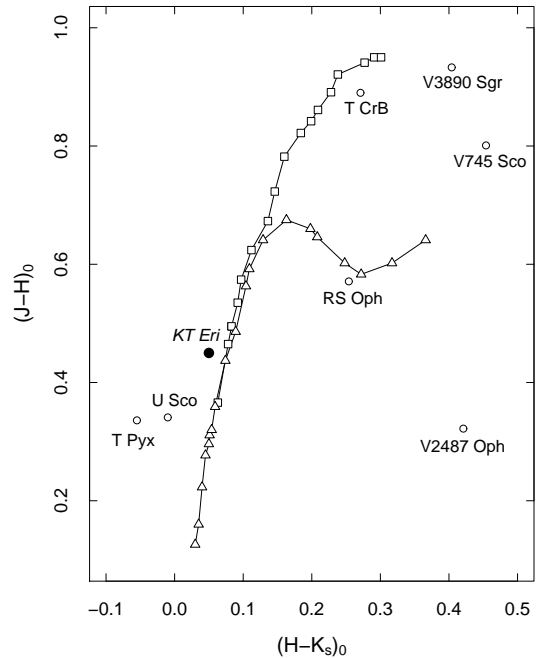}
      \end{center}
     \caption{The color--color diagram of $(J-H)_0$ vs $(H-K_s)_0$. 
              The data of KT Eri, V2487 Oph and RS Oph are from 2MASS Point Source Catalog.
              These infrared data of U Sco (\cite{han1985}), V745 Sco, V3890 Sgr, 
              T CrB and T Pyx (\cite{har1993}) are transformed to the system of 
              2MASS (\cite{carp2001}). 
              The lines connecting the squares and triangle are the 
              colors of giant stars from G0 III to M7 III and main sequence stars from 
              F0 V to M6 V, respectively (\cite{bb1988}).        
              }
  \end{figure}

\subsection{Recurrency of outbursts}
  Temporal behavior of KT Eri resembles some of the RNe, 
for example; V745 Sco and V3890 Sgr (e.g., \cite{wil1991}, \cite{sch2010}), 
but is not similar to T Pyx or IM Nor (e.g., \cite{sch2010}, \cite{kat2002}). 
As is known, RNe are divided into three subclasses (\cite{war1995}, 2008).
From the obtained expansion velocity, KT Eri is thought to have intermediate properties 
between T CrB subclass and U Sco subclass.
From our spectral observations, KT Eri is supposed to be a spectral class of He/N 
classical nova. However, as is pointed out by \citet{kat2004}, 
such a spectral classification does not necessarily exclude the possibility of RN. 

  On the other hand, RNe have generally a higher accretion rate than CNe 
due to massive WD (e.g., Hachisu \& Kato 1999, 2001a, 2001b, 
\cite{hach2002}; \cite{hell2001} for a review). 
Applying the WD mass-decline rate relation suggested by \citet{kathach1994} to this case, 
KT Eri is to have a massive WD. 
If this is true, KT Eri is considered to be a RN ($T_R > 100$ yr though; see \cite{jur2012}).

\section{Summary}
   Table 6 summarizes the parameter of this star which are revealed in this study.
KT Eri is an exotic nova which appeared on the high galactic latitude and 
opposite side from galactic center.
Spectral class and speed class are He/N and \textit{very fast}, respectively.
The distance is approximately 7 kpc and  the galactic height  $z \sim 4$ kpc. 
Hence, KT Eri is thought to be located outside of the galactic disk.
Accordingly, it is plausible that secondary star is a giant.

  \begin{table}
  \caption{The parameter of KT Eri in this study. 
           }
  \begin{center}
 \begin{tabular}{cr} \hline
   spectral class           & He/N                                           \\
     speed class            & very fast                                      \\
     $t_2$                  & $6.2 \pm 0.3$ d                                \\
     $t_3$                  & $14.3 \pm 0.7$ d                               \\
     $t_0$                  & 2009 Nov. $14.4 \pm 0.2$ UT                    \\
     $m_{V}$(max)           & 5.4 mag                                        \\
     $M_{V}$(max)           & $\sim -9$ mag                                  \\
     $M_{V}$(min)           & $\sim 0.4$ mag                                 \\
     $d$                    & $\sim 7$ kpc                                   \\
     $z$                    & $\sim 4$ kpc                                   \\
      
 \hline
 \end{tabular}
 \end{center}
 \end{table}

 \vspace{20pt}

 The authors are grateful for members of OUS observational team (N. Kunitomi, M. Nose and R. Takagi).
 We would like to express gratitude to ASAS, Pi of the sky, VSOLJ and 2MASS for their useful data.
 We also thank Dr. K. Ayani (BAO director) for his observational support and Dr. N. Fukuda (OUS) for 
 his kind advice.


\begin{thebibliography}{}

 \bibitem[Bessell \& Brett(1988)]{bb1988}
  Bessell, M. S. \& Brett, J. M. 1988, \pasp, 100, 1134
 
 \bibitem[Bode et al.(2010)]{Bod2010}
  Bode, M. F. et al. 2010, The Astronomer's Telegram, 2392, 1

 \bibitem[Capaccioli et al.(1989)]{cap1989}
  Capaccioli, M., della Valle, M., Rosino, L, \& D'Onofrio, M. 1989, \aj, 97, 1622
  
 \bibitem[Cardelli et al.(1989)]{car1989}
  Cardelli, J. A., Clayton, G. C. \& Mathis, J. S. 1989, \apj, 345, 245

 \bibitem[Carpenter(2001)]{carp2001}
  Carpenter, J. M. 2001, \aj, 121, 2851

 \bibitem[Cohen(1988)]{coh1988}
  Cohen, J. G. 1988, in ASP Conf. Ser. Vol. 4, ed. van den Bergh, S. \& Pritchet, C. J., 114

 \bibitem[Darnley et al.(2012)]{darn2012}
  Darnley, M. J. et al. 2012, \apj, 746, 61


 \bibitem[Della Valle et al.(1992)]{del1992}
  Della Valle, M., Bianchini, A., Livio, M. \& Orio, M. 1992, \aap, 266, 232
  
 \bibitem[Della Valle \& Livio(1995)]{del1995}
  Della Valle, M. \& Livio, M. 1995, \apj, 452, 704
  
 \bibitem[Della Valle \& Livio(1998)]{del1998}
  Della Valle, M. \& Livio, M. 1998, \apj, 506, 818
  
 \bibitem[Diaz et al.(2010)]{diaz2010}
  Diaz, M. P. et al. 2010, \aj, 140, 1860
 
 \bibitem[Downes \& Duerbeck(2000)]{dow2000}
  Downes, R. A. \& Duerbeck, H. W. 2000, \aj, 120, 2007 

 \bibitem[Downes et al.(2005)]{dow2005}
  Downes, R. A., Webbink, R. F., Shara, M. M., Ritter, H., Kolb, U. \& Duerbeck, H. W. 2005, VizieR Online Data Catalog, 5123



 \bibitem[Duerbeck(1981)]{due1981}
  Duerbeck, H. W. 1981, \pasp, 93, 165
 
 \bibitem[Gallagher \& Starrfield(1978)]{gal1978}
  Gallagher, J. S. \& Starrfield, S. 1978, \araa, 16, 171
 
 \bibitem[Hachisu \& Kato(1999)]{hachkato1999}
  Hachisu, I. \& Kato, M. 1999, \apj, 517L, 47
 
 \bibitem[Hachisu \& Kato(2001)]{hachkato2001a}
  Hachisu, I. \& Kato, M. 2001a, \apj, 558, 323
 
 \bibitem[Hachisu \& Kato(2001)]{hachkato2001b}
  Hachisu, I. \& Kato, M. 2001b, \apj, 553L, 161
 
 \bibitem[Hachisu et al.(2002)]{hach2002}
  Hachisu, I., Kato, M., Kato, T. \& Matsumoto, K. 2002, ASPC, 261, 629
 
 \bibitem[Hanes(1985)]{han1985}
  Hanes, D. A. 1985, \mnras, 213, 443
 
 \bibitem[Harrison et al.(1993)]{har1993}
  Harrison, T. E., Johnson, J. J. \& Spyromilio, J. 1993, \aj, 105, 320
 
 \bibitem[Hellier(2001)]{hell2001}
  Hellier, C. 2001, in Cataclysmic Variable Stars (Berlin: Springer)
  
 \bibitem[Hounsell et al.(2010)]{hou2010}
  Hounsell, R. et al. 2010, \apj, 724, 480


 
 \bibitem[Jurdana-{\v S}epi{\'c} et al.(2012)]{jur2012}
  {Jurdana-{\v S}epi{\'c}}, R., Ribeiro, V.~A.~R.~M., Darnley, M.~J., Munari, U. \and Bode, M.~F. 
  2012, \aap, 537, 34
 
 \bibitem[Kato \& Hachisu(1994)]{kathach1994}
  Kato, M. \& Hachisu, I. 1994, \apj, 437, 802
 
 \bibitem[Kato et al.(2002)]{kat2002}
  Kato, T., Yamaoka, H., Liller, W. \& Monard, B. 2002, \aap, 391L, 7
 
 \bibitem[Kato, Yamaoka \& Kiyota(2004)]{kat2004}
  Kato, T., Yamaoka, H. \& Kiyota, S. 2004, \pasj, 56, 83
 
 \bibitem[Kraft(1958)]{kra1958}
  Kraft, R. P. 1958, \pasp, 70, 598

 \bibitem[Munari \& Zwitter(1997)]{mun1997}
  Munari, U. \& Zwitter, T. 1997, \aap, 318, 269

 \bibitem[Ootsuki et al.(2009)]{Oot2009}
  Ootsuki, I. et al. 2009, \iaucirc, 9098, 2
 
 \bibitem[O'Brien et al.(2010)]{OBr2010}
  O'Brien, T. J. et al. 2010, The Astronomer's Telegram, 2434, 1
    
 \bibitem[Payne-Gaposchkin(1957)]{pay1957}
  Payne-Gaposchkin, C. 1957, in Galactic Novae (Amsterdam: North-Holland P. C.)
  

 \bibitem[Pojma{\'n}ski(2002)]{asas}
  Pojma{\'n}ski, G. 2002, AcA, 52, 397

 \bibitem[Ragan et al.(2009)]{rag2009}
  Ragan, E. et al. 2009, The Astronomer's Telegram, 2327, 1

 \bibitem[Schaefer(2010)]{sch2010} 
  Schaefer, B. E. 2010, \apj, 187, 275 
 
 \bibitem[Sekiguchi et al.(1988)]{Sek1988}
  Sekiguchi, K. et al. 1988, \mnras, 234, 281
 
 \bibitem[Sekiguchi et al.(1989)]{Sek1989}
  Sekiguchi, K. et al. 1989, \mnras, 236, 611
 
 \bibitem[Starrfield(1989)]{sta1989}
  Starrfield, S. 1989, in Classical Novae, ed. Bode, M. F. \& Evans, A. (New York: Wiley \& Sons Ltd.), ch. 3
 
 \bibitem[Yamaoka et al.(2009)]{yam2009}
  Yamaoka et al. 2009, \iaucirc, 9098, 1

 \bibitem[Walker(1954)]{wal1954}
  Walker, M. F. 1954, \pasp, 66, 230
 


 \bibitem[Warner(1987)]{war1987}
  Warner, B. 1987, \mnras, 227, 23
 
 \bibitem[Warner(1995)]{war1995}
  Warner, B. 1995, in Cataclysmic Variable Stars (New York: Cambridge University Press)

 \bibitem[Warner(2008)]{war2008} 
  Warner, B. 2008, in Classical Novae, ed. Bode, M. F. \& Evans, A. (New York: Cambridge University Press)

 \bibitem[Williams(1990)]{wil1990}
  Williams, R. E. 1990, in Physics of Classical Novae, ed. Cassatella, A. \& Viotti, R. (Berlin: Springer-Verlag), 215
  
 \bibitem[Williams(1991)]{wil1991}
  Williams, R. E. et al. 1991, \apj, 376, 721

 \bibitem[Williams(1992)]{wil1992}
  Williams, R. E. 1992, \aj, 104, 725



 \bibitem[Williams, {Phillips} \& {Hamuy}(1994)]{wil1994}
  Williams, R. E., Phillips, M. M. \& Hamuy, M. 1994, \apjs, 90, 297


\end{thebibliography}
\end{document}